\documentclass{aa}  

\usepackage{xcolor}
\usepackage{graphicx}
\usepackage{subcaption}
\usepackage{color}
\usepackage{txfonts}
\usepackage{tikz}
\usetikzlibrary{positioning}
\usepackage{amssymb}
\usepackage{graphicx}
\usepackage{amsmath}
\usepackage{pict2e}
\usepackage{multirow}
\usepackage{todonotes}
\usepackage{todonotes}

\def\ML{\textrm{machine learning}}

\begin{document} 

\title{Supervised Neural Networks for\\ Helioseismic  Ring-Diagram Inversions}
\titlerunning{Supervised Neural Networks for\\ Helioseismic  Ring-Diagram Inversions}
   \author{Rasha Alshehhi\inst{1}
          \and Chris S. Hanson\inst{2,1}
          \and Laurent Gizon\inst{2,3,1}
          \and Shravan Hanasoge\inst{4,1}
          }

\institute{
\inst{1} Center for Space Science, NYUAD Institute, New York University Abu Dhabi, UAE \\
\inst{2} Max-Planck-Institut f\"ur Sonnensystemforschung, Justus-von-Liebig-Weg 3, 37077 G{\"o}ttingen,   Germany \\
\inst{3} Institut f\"{u}r Astrophysik, Georg-August-Universit\"{a}t G\"{o}ttingen, Friedrich-Hund-Platz 1, 37077 G{\"o}ttingen,   Germany \\
\inst{4} Tata Institute of Fundamental Research, Homi Bhabha Road, Mumbai 400005, India\\
\email{ra130@nyu.edu,hanson@mps.mpg.de}              
              }

   \date{Received \today; accepted  XXX}

 
  \abstract
   {The inversion of ring fit parameters to obtain subsurface flow maps in ring-diagram analysis for 8 years of SDO observations is computationally expensive, requiring $\sim$3200 CPU hours.}
   {In this paper we apply machine learning techniques to the inversion step of the ring-diagram pipeline in order to speed up the calculations. Specifically, we train a predictor for subsurface flows using the mode fit parameters and the previous inversion results, to replace future inversion requirements.}
   { We utilize Artificial Neural Networks as a supervised learning method for predicting the flows in $15^\circ$ ring tiles. We discuss each step of the proposed method to determine the optimal approach. In order to demonstrate that the \ML{} results still contain the subtle signatures key to local helioseismic studies, we use the \ML{} results to study the recently discovered solar equatorial Rossby waves.  }
   {The Artificial Neural Network is computationally efficient, able to make future flow predictions of an entire Carrington rotation in a matter of seconds, which is much faster than the current $\sim31$~CPU hours. Initial training of the networks requires $\sim${3} CPU hours. The trained Artificial Neural Network can achieve a root mean-square error equal to approximately half that reported for the velocity inversions, demonstrating the accuracy of the \ML{} (and perhaps the overestimation of the original errors from the ring-diagram pipeline). We find the signature of equatorial  Rossby waves in the \ML{} flows covering six years of data, demonstrating that small-amplitude signals are maintained. The recovery of Rossby waves in the \ML{} flow maps can be achieved with only one Carrington rotation ($27.275$~days) of training data. }
   {We have shown that machine learning can be applied to, and perform more efficiently than the current ring-diagram inversion. The computation burden of the \ML{} includes {3} CPU hours for initial training, then around $10^{-4}$ CPU hours for future predictions. 
   }
   \keywords{Sun: helioseismology -- Sun: oscillations -- Sun: interior -- Methods: numerical -- Methods: data analysis             }

   \maketitle
%


\section{Motivation: Speeding up ring-diagram inversions}

Local helioseismology seeks to image the subsurface flows utilizing the complex wave field observed at the surface \citep[see review by ][]{gizon_birch_2005}. The procedure of imaging the subsurface flow fields from the Dopplershift images of the surface is summarized as follows; the projection and tracking of large Dopplergram times series, transformation into Fourier space, analysis of perturbations in the acoustic power spectrum (local frequency shifts) and inversions. The refinement of large data sets into the significantly smaller flow maps is computationally expensive, and has to be repeated for all future observations. The field of machine learning seeks to develop data driven models that, given sufficient training samples, will predict future observations, without the need for the aforementioned procedure. With over 20 years of space-based observations, the field of local helioseismology now possesses large amounts of data that can be utilized by machine learning algorithms to improve existing techniques, or find relationships previously unknown to the field. 

The field of machine learning grew out of the work to build artificial intelligence. The application of machine learning is broad in both scientific research and industries that analyze `big data', with some impressive results \citep[e.g.][]{pearson_etal_2018}. However, the field of local helioseismology has thus far not utilized this technique, despite the advantages it could provide given  the large amounts of data available. {However,} some work has been done in using deep learning for multi-height local correlation tracking near intergranular lanes \citep{asensioramos_etal_2017}.  

A widely used technique in local helioseismology is ring-diagram analysis \citep[see][for detailed review]{antia_basu_2007}. First presented by \citet{hill_1988}, the ring diagram technique analyzes slices (at constant frequency $\omega$) of the 3D power spectrum $P(\omega,k_x,k_y)$ of an observed (tracked and projected) patch of the solar surface $D(t,x,y)$. The cross-section of the power spectrum reveals rings corresponding to the acoustic normal modes of the Sun (f- and p-modes). In the absence of flows these rings are symmetric in $k_x$ and $k_y$. However, the presence of flows in the zonal ($x$) or meridional ($y$) directions breaks symmetry of these rings, manifesting as ellipsoids. Acoustic modes traveling with or against the direction of flow experience increases or decreases in travel time, resulting in changes to the phase speed. The frequency shift of a ring is then considered as a horizontal average of the flows within the observed patch. Each mode (ring) is then fit, and the mode-fit parameters $U_x$ and $U_y$ determined, revealing the `flow' required to produce the shifts in each mode (ring). The true subsurface flow field is then considered as a linear combination of the mode-fits. 
In order to determine the flow field, an inversion is performed using the mode-fit parameters $U_x$ and $U_y$ and the sensitivity kernels $\mathcal{K}^{n\ell}(z)$ that relate frequency shifts of a mode of radial order $n$ and harmonic degree $\ell$ to the horizontal velocity components $u_x$ and $u_y$ at height $z$ in the interior (by convention, $z$ is negative in the interior and zero at the photosphere):
\begin{equation}
U^{n\ell}_x = \int u_x(z)\, \mathcal{K}^{n\ell}(z) \, \textrm{d}z, \quad  U^{n\ell}_y = \int u_y(z)\, \mathcal{K}^{n\ell}(z) \, \textrm{d}z ,
\end{equation}
where $\mathcal{K}^{n\ell}$ is derived from solar models. The solution to the linear inversion is then a combination of the mode fits and coefficients $c^{n\ell}$ that give maximum sensitivity at a target height $z_t$,
\begin{eqnarray}
u_x(z_t) = \sum\limits_{n,l} c^{n\ell} (z_t) \, U_x^{n\ell}, \quad
u_y(z_t) = \sum\limits_{n,l} c^{n\ell} (z_t) \, U_y^{n\ell}.
\end{eqnarray}
The ring-diagram pipelines that derive these coefficients \citep{bogart_etal_2011a,bogart_etal_2011b} are computationally expensive, requiring 16~s per tile or $\sim3200$ CPU hours for the entire SDO data set.

Our endeavor is to utilize the large data sets available in the pipeline, to improve the ring diagram procedure by utilizing deep network. Specifically, we seek (through machine learning) to determine the complex mapping from the raw Doppler time series to the flows. 
In this study we present initial results in performing the inversion without the need for any inversions or kernels, utilizing \ML{} techniques.

In section~\ref{sec.proposed_method} we present the data to be used and the proposed \ML{} technique. Section~\ref{sec.performance} will examine the performance of the proposed method with a number of \ML{} techniques in both accuracy and computational burden. Section~\ref{sec.rossby} examines a case study of the effect of \ML{} on equatorial Rossby waves. Discussions and conclusions are given in Section~\ref{sec.conclusions}.


\section{Proposed Method}\label{sec.proposed_method}
Machine learning studies can be divided into two branches, unsupervised and supervised learning. This study will be of the latter kind, in which we know the flow corresponding to the mode fits (from the current pipeline) and thus train an estimator to predict flow values given a new set of mode fits. While no study has directly examined the accuracy of ring-diagram analysis, the results of a number of studies have remained consistent with other measurement techniques \citep[e.g.][]{giles_etal_1997,schou_bogart_1998}. 
{It is possible that the existing pipeline has systematic errors that affect the inversion results. Any supervised learning method will inherit these problems, as this is the weakness of data driven models. If problems were found and resolved in the pipeline, any trained machine learning models would have to be retrained for the correct flows, although this will not invalidate the results of this study.}
The proposed supervised method of this study comprises two main components: preprocessing and applying an ANN for regression, both of which are described in detail in the following sections.

For clarity in the terminology used here, we define the following. Input data will consist of a large number $N_{obs}$ of tiles, each with a number of mode-fits/features $N_{\textrm{features}}$ identified in the ring pipeline. The output data consists of $N_{obs}$ flow values, corresponding to each input tile, for 31 depths.

\subsection{Extraction of features from pipeline }
The ring diagram pipeline \citep{bogart_etal_2011a,bogart_etal_2011b} developed for use on the high resolution, high cadence data of the Helioseismic and Magnetic Imager \citep[HMI,][]{schou_etal_2012b}, has provided unprecedented analysis of the Sun's subsurface flows. The data for each step in the pipeline is available on the NetDRMS~\footnote{http://jsoc.stanford.edu/netdrms/} database, and for this study we utilize the mode fits $U_x$ and $U_y$ and the inverted flows $u_x$ and $u_y$. Due to the statistical nature of the \ML{}, we ignore the derived error values of the fits in the training. {However, in Sec.~\ref{sec.results} we will show the {effect} realization noise has on the \ML{} predictions.}

The SDO program has been running since 2010, and has observed over 100 Carrington rotations (CRs). For this study we make use of the data from CR2097 to CR2201, which covers eight years. {Each Carrington rotation consists of a maximum of 6816 tiles, but some rotations have less coverage. In total, over the 104 rotations, there are 709734 {inversion results} available in the pipeline.} Additionally, we focus this study on the $15^\circ$ maps, which upon inversion probes depths down to $20.88$~Mm below the photosphere.

\begin{figure}
\includegraphics[width=\linewidth]{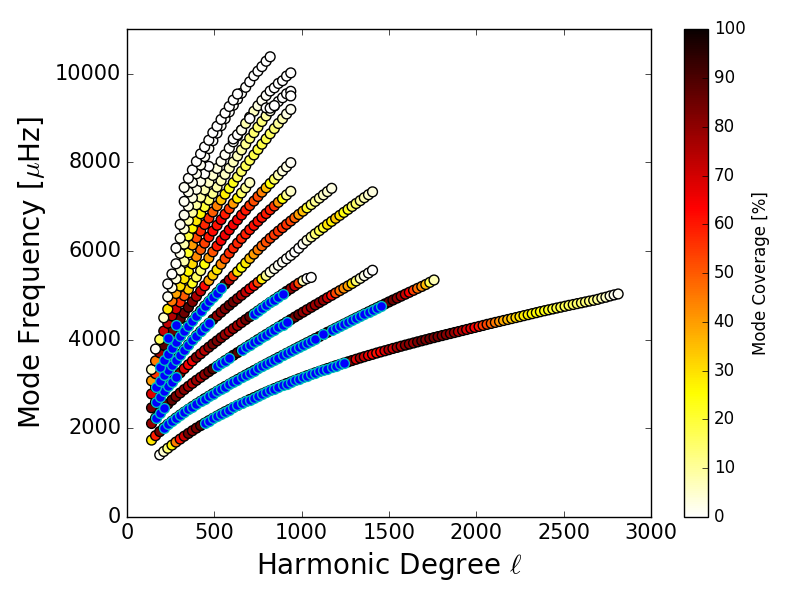}
\caption{The mode coverage of each uniquely identified $[n,\ell]$ mode in the mode-fits pipeline from CR2097 to CR2201. The modes highlighted with blue appear in more than 90\% of the tiles and thus are used in this study. All other modes are neglected.}
\label{fig.modeCoverage}
\end{figure}

Each tile has a number of mode fits that have been detected by the pipeline. From tile to tile the presence of these modes can vary, sometimes detected, other times overlooked. For this study each unique mode, with radial order $n$ and harmonic degree $\ell$, is considered an independent feature. The presence of a single mode in all tiles is called the mode coverage. In order to avoid bias from missing modes, we reduce the number of features by applying strict mode coverage requirements. Specifically, for a single mode, if it is detected in less than 90\% of the tiles, then it is neglected. This significantly reduces the number of features in the \ML{}, and minimizes any bias (to zero) we have from missing data. Upon application of this mode coverage requirement, 152 separate modes remain. Figure~\ref{fig.modeCoverage} shows the mode coverage for all modes detected in the pipeline, as well as the modes selected for this \ML{} study.

In summary the dataset (for each horizontal component $x$ and $y$ of the flow) consists of 709734 tiles (samples) with 152 features (modes) each specifying the frequency shifts of acoustic waves due to flows. The corresponding target consists of 709734 flow vectors for each of the 30 depths from a depth of 0.696~Mm to 20.88~Mm. 

\subsection{Preprocessing of input features}
The goal of preprocessing the input features is to produce more effective features which show high contrasts. Typically, this involves three steps: interpolation to fill missing data, normalization of the input features, and reduction of the number of input features.

Previously we neglected modes that appeared in less than 90\% of the tiles. However, many of the tiles still do not have complete mode coverage. The problem of how to handle missing data is well known in machine learning. {In statistics and machine learning literature, the replacement of missing data is known as data imputation.} A number of solutions exist in completing the missing data \citep[e.g.][]{little_rubin_1987}. However, for this study we utilize the simple solution of replacing the missing values by the mean of the mode. 

The next stage is the normalization of each feature, which will bring all features into a similar dynamic range. The need for this is driven by the fact that many machine learning techniques will be dominated by features with a large range, while narrow ranged features can be ignored. To avoid this, normalization of each feature is performed in order to bring all features into the same range, and thus remove any preference for a specific feature. There are a number of different approaches that can be taken to achieve this (e.g., minimum-maximum feature scaling, standard score, student's t-statistic and studentized residual), and we recommend the reader to read \citet{Juszczak:2002} for an in-depth explanation of each approach. For this study we use standard score. The transformation of a vector of  frequency shift measurements $\boldsymbol{X}$ to the new normalized feature $\tilde{X}_i$ of observation $i$ is as follows:
\begin{equation}
\tilde{X}_i = \frac{{X}_i - \mu}{\sigma},
\label{eq:scale}
\end{equation}
where $\mu$ is the mean of the elements of $\boldsymbol{X}$ and $\sigma$ the standard deviation. The new features $\tilde{\boldsymbol{X}}$ will have a zero mean with unit variance.

The final step in the preprocessing of this study is reduction of the 152 features (modes) to a smaller number, in order to ease computational burden. 
Typically, the processing of features is done through either feature selection; i.e. choosing only a subsample of modes or feature reduction; i.e. new feature space is generated from original modes. 
\citep[see][Chapter 6]{alpaydin_2010}. By limiting our study to those 152 features with sufficient mode coverage, we have already performed feature selection. However, the remaining number of modes is still quite high and we seek to further reduce this through feature reduction. Here, feature reduction is achieved through Canonical Correlation Analysis \citep[CCA,][]{hotelling_1936,cca}. 

{
The CCA seeks linear combinations of the input data $X$ and output data $Y$ (flow velocities), which have a maximum correlation with each other. Specifically, we seek Canonical Correlation Vectors $a_i$ and $b_i$ that maximize the correlation
\begin{equation}\label{eq.corrCCA}
\rho(a_i,b_i) = \textrm{Corr}(a_i^TX,b_i^TY).
\end{equation}
Following the method outlined by \citet{haerdle_simar_2007}, it can be shown that $a_i$ and $b_i$ are related to the covariance matrices $\Sigma_{XX} = \textrm{Cov}(X,X)$ and $\Sigma_{YY} = \textrm{Cov}(Y,Y)$  through
\begin{equation}
\begin{aligned}
a_i &= \Sigma_{XX}^{-1/2}\mathcal{U}_i, \\
b_i &= \Sigma_{YY}^{-1/2}\mathcal{V}_i,
\end{aligned}
\end{equation}
where $\mathcal{U}_i$ and $\mathcal{V}_i$ are the $i$\textsuperscript{th} left and right singular vectors from the following singular value decomposition (SVD): 
\begin{equation}
\begin{aligned}
\textrm{SVD}\left(\Sigma_{XX}^{-1/2}\Sigma_{XY}\Sigma_{YY}^{-1/2}\right) = [ \cdots \mathcal{U}_i \cdots ] \Lambda
[\cdots \mathcal{V}_j \cdots  ]^T, 
\end{aligned}
\end{equation}
with $\Lambda$ the diagonal matrix of singular values {and $\Sigma_{XY} = \textrm{Cov}(X,Y)$}.
It remains to be determined how many Canonical correlation vectors are required to capture the relationship between $X$ and $Y$. In section~\ref{sec.results} we will show that upon application of CCA the number of features in the input data reduces from 152 to 1{, for each depth and flow component}. 

{Figure~\ref{fig:cca_analysis_coef} shows the coefficients of the modes for feature reduction, computed through CCA, for two target depths. Over plotting the phase speed corresponding to modes with a lower turning point at the target depth, shows that the CCA gives more weight to modes that are sensitive to horizontal flows at the target depth. Thus, while we have reduced the 152 features to 1 (for each depth and flow component) the sensitivity to horizontal flows at the target depth is maintained.}

{Figure~\ref{fig:cca_analysis_kernel} compares the averaging kernel computed for tile hmi.rdVtrack\_fd15[2196][240][240][0][0] with one built with the coefficients derived through the CCA. While the CCA finds the coefficients that maximize the correlations between mode-fits and flows for all tiles, the results show that the kernels compare reasonably well (despite no prior requirement on depth localization). We note that the CCA kernels are more sensitive to the surface for deep $z_t$, indicating that the CCA may find some depth correlations. The exact correlation of flows with depth is beyond the scope of this study, but is worth future investigation.}

\begin{figure*}
    \begin{subfigure}[b]{0.55\linewidth}
    \centering
    \includegraphics[width=\linewidth]{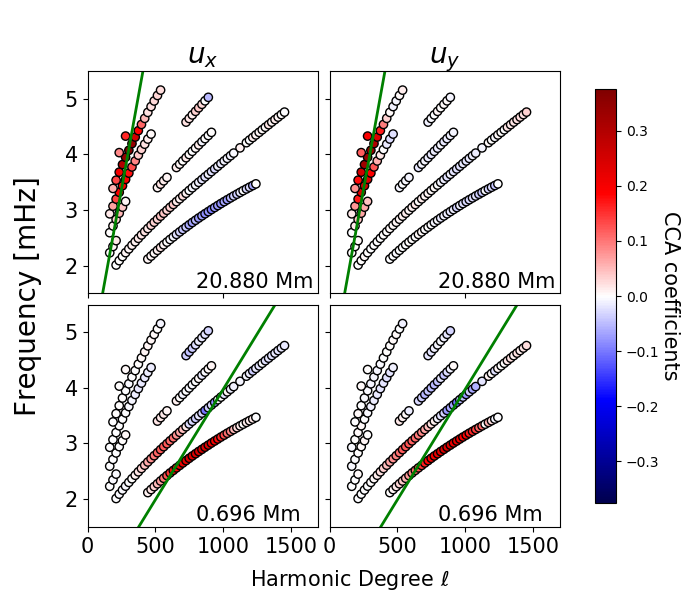}
    \caption{}
    \label{fig:cca_analysis_coef}
    \end{subfigure}
    \begin{subfigure}[b]{0.44\linewidth}
    \centering
    \includegraphics[width=\linewidth]{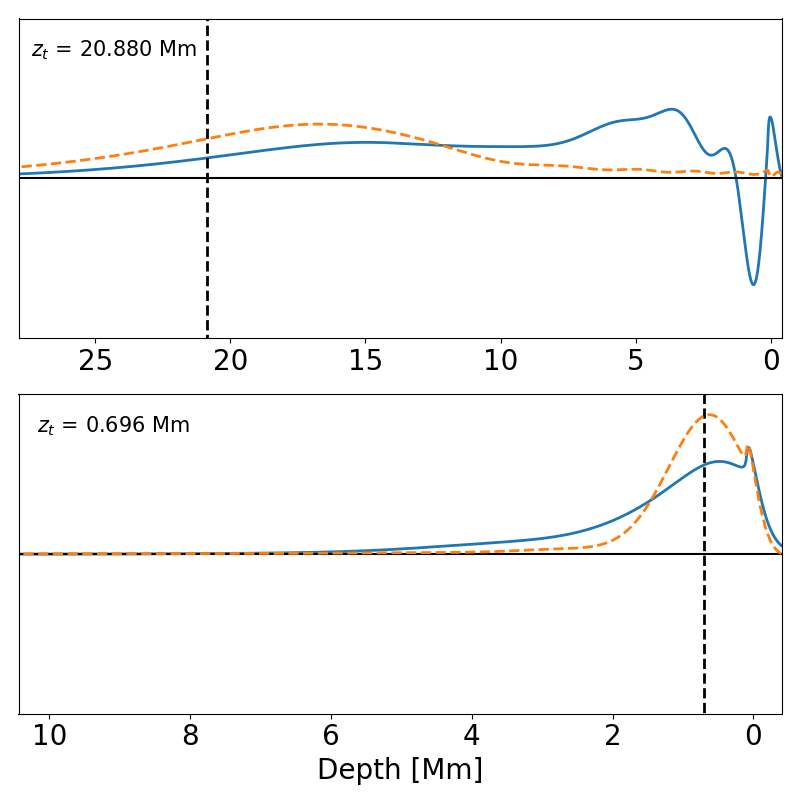}
    \caption{}
    \label{fig:cca_analysis_kernel}
    \end{subfigure}
    \caption{{Panel (a):The coefficients computed using cross-correlation analysis for $u_x$ (left) and $u_y$ (right) to reduce the 152 unique modes to a single feature for each depth and flow component. Here we show the coefficients for two depths, 20.88~Mm (top) and 0.696~Mm (bottom). The CCA computes the ideal weights for the original data, given correlations with the output flows at a single depth. The phase speed corresponding to a lower turning points at the target depth is also shown (green line). Modes close to this phase speed are given greater weight by the CCA.
    Panel (b): The averaging kernel build using the inversion pipeline (orange dashed) for a single tile, and an equivalent kernel built using the CCA coefficients derived using all tiles (blue line).}}
    
    \label{fig:cca_analysis}
\end{figure*}

A schematic of the entire preprocessing pipeline is shown in Fig~\ref{fig:preprocess}.

\begin{figure*}[!htb]
\centering
\tikzset{%
  every neuron/.style={
    rectangle,
    draw,
    minimum size=1cm
  },
  neuron missing/.style={
    draw=none, 
    scale=3,
    text height=0.2cm,
    execute at begin node=\color{black}$\vdots$
  },
}
\begin{tikzpicture}[x=1.3cm, y=1.3cm, >=stealth]

\node  (input) at (0,0) 
{\begin{tabular}{c}
$U_x$ or $U_y$\\
(mode fits)
\end{tabular}};

\node [every neuron/.try,label=below:$\geq90\%$ coverage] (mode) at (2,0) 
  {\begin{tabular}{c}
  Mode\\ Selection
  \end{tabular}};
  
\node [every neuron/.try,label=below:Mean Filling] (impute) at (4,0) {Imputation};

\node [every neuron/.try,label=below:Standardization] (scaling) at (6,0) 
{\begin{tabular}{c}
Feature\\ Scaling
\end{tabular}};

\node [every neuron/.try,label=below:CCA (1 depth)] (reduction) at (8,0) {
  \begin{tabular}{c}
  Feature\\ Reduction
  \end{tabular}};

\node (output) at (11,0) 
{\begin{tabular}{c}
$\tilde{X}(z_t)$\\
(Reduced mode-fits)
\end{tabular}};

\draw [->] (input) -- (mode);
\draw [->] (mode) -- (impute);
\draw [->] (impute) -- (scaling);
\draw [->] (scaling) -- (reduction);
\draw [->] (reduction) -- (output);

\end{tikzpicture}
\caption{The preprocessing pipeline proposed by this study for preparing the mode fit parameters for the \ML{}. The pipeline must be performed for both $U_x$ and $U_y$, while the {CCA must be performed for each depth}. The output $\tilde{X}(z_t)$ is then the input for the ANN (Fig.~\ref{fig:nn}).}
\label{fig:preprocess}
\end{figure*}
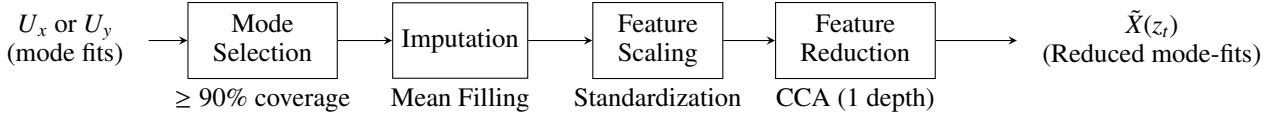

}

\subsection{Neural Networks}
The Artificial Neural Network (ANN) is one of the most common supervised machine learning methods with a wide range of literature \citep[e.g.][]{Hand:2001,Haykin:2009,Bishop:2006}. While many other methods exist in \ML{}, we have found that the ANN performs best for this study (see Section~\ref{sec.results}), and thus will explain here in detail the ANNs we utilize. For an overview of other methods see \citet{alpaydin_2010}. 
One advantage of the ANN is that it can solve non-linear problems, which arise in this study due to different modes sets used in each tile's inversion (dependent on noise and disk position etc.) that directly feed into the inversion results. 

\begin{figure*}
\centering
\tikzset{%
  every neuron/.style={
    circle,
    draw,
    minimum size=1cm
  },
  neuron missing/.style={
    draw=none, 
    scale=3,
    text height=0.2cm,
    execute at begin node=\color{black}$\vdots$
  },
}
\begin{tikzpicture}[x=1.3cm, y=1.3cm, >=stealth]

\foreach \m/\l [count=\y] in {1}
  \node [every neuron/.try, neuron \m/.try] (input-\m) at (0,2.5-3*\y){};

\foreach \m [count=\y] in {1,2,missing,3}
  \node [every neuron/.try, neuron \m/.try ] (hidden-\m) at (2,2-\y*1.25) {};


\foreach \m [count=\y] in {1}
  \node [every neuron/.try, neuron \m/.try ] (output-\m) at (4,1.5-\y*2) {};

\foreach \l [count=\i] in {1}
  \draw [<-] (input-\i) -- ++(-1,0)
    node [above, midway] {$\tilde{X}(z_t)$};

\foreach \l [count=\i] in {1,2,M}
  \node [above] at (hidden-\i.north) {$H^1_\l$};


\foreach \l [count=\i] in {1}
  \draw [->] (output-\i) -- ++(1,0)
    node [above, midway] {$\hat{y}(z_t$)};

\foreach \i in {1}
  \foreach \j in {1,...,3}
    \draw [->] (input-\i) -- (hidden-\j);


\foreach \i in {1,...,3}
  \foreach \j in {1}
    \draw [->] (hidden-\i) -- (output-\j);

\foreach \l [count=\x from 0] in {Input, Hidden, Output}
  \node [align=center, above] at (\x*2,2) {\l \\ layer};

\end{tikzpicture}
\caption{The general structure of the Multi-Layer Perceptron. The input layer consists of 1 passive node, which is the output of the preprocessing pipeline, that relays the values directly to the hidden layer. The hidden layers $H^1$ consists of $M$  {non-linear} active neurons, that modify their input values and produce a signal to be fed into the next layer. The output layer then transforms its input values (from the hidden layer) into the flows values at target height $z_t$.}
\label{fig:nn}
\end{figure*}
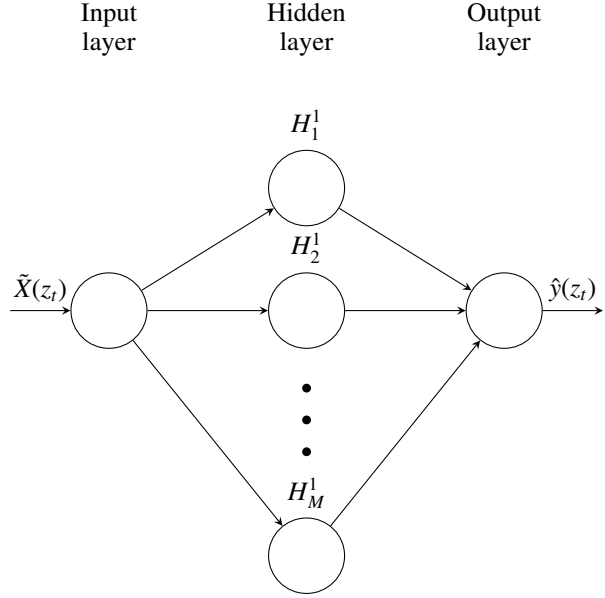

In this work a Multi-Layer Perceptron (MLP) neural network \citep[e.g.][]{haykin_1998} is used (see Fig.~\ref{fig:nn} for example). This class of the ANN is known as a feed-forward ANN, where each layer consists of multiple neurons (or activation functions) acting as fundamental computation units. Connectivity is unidirectional from neurons in one layer to neurons in the next layer such that the outputs of a neuron in a layer serve as input to neurons in the following layer. The degree to which each neuron is activated is specified by the weight of the neuron.

The MLP utilizes supervised learning in order to determine the correct weights for each neuron. This algorithm proceeds in two phases: forward and backward propagation. The network is initialized with random values for the weights. The forward propagation then runs the input through the network, generating outputs based on the initial layer weights. In the backward propagation (BP), the errors between the ANN outputs and actual values (flows) are computed. Using this error, the weights of each activation function are then updated (through stochastic gradient descent) in order to minimize the output errors. The BP algorithm is performed in mini-batches {(200 samples)} of the total dataset, with several passes over the entire set until convergence is achieved. {Unlike classic stochastic gradient descent which updates the weights after every sample pass through the ANN, mini-batches settle on the minimum better because they are less subjected to noise.} On average, each iteration will improve the weights, minimizing the difference between {the predicted output and pipeline output}. 

The specific algorithm for the forward and back propagation is as follows. For {ring-diagram} tile (sample) $k$ in mini-batch $n$, let $y_{j,k}^l(n)$ be the output of neuron $j$ in the layer $l$ (which consists of $m_l$ neurons) and $w_{ji}^{l}(n) \in \mathbb{R}$ be the weight applied to the output of neuron $i$ in the previous layer $l-1$ fed into neuron $j$ in layer $l$. The previous layer ($l-1$) consists of $m_{l-1}$ neurons. In the forward propagation, the weights are fixed and the outputs are computed on a neuron by neuron basis:
\begin{equation} \label{eq.neuronoutput}
y_{j,k}^{l}(n)=  \varphi(\upsilon_{j,k}^l(n)),
\end{equation}
where 
\begin{equation}
\upsilon_{j,k}^l(n):=  b^l_j(n) + \sum\limits_{i=1}^{m_l}w_{ji}^{l}(n)y_{i,k}^{l-1}(n)
\end{equation}
and the function $\varphi$ refers to the activation function and $b_j^l(n)$ is the bias. In this work, the activation function in the hidden layers is the Rectifier Linear Unit (ReLU) function  while {the identity function is used for the output layer} 
\begin{equation}
\varphi(\upsilon_{j,k}^l(n)) = 
\left\{
\begin{array}{ll}
\textrm{max}(0,\upsilon_{j,k}^l(n))
& l = \text{hidden} ,
\\
\upsilon_{j,k}^l(n) & l = \text{output}. 
\end{array}
\right.
\label{eq:relu}
\end{equation}
{Our choice in the ReLU function for the hidden layers is due to it's improved convergence over other activation functions and suffers less from the vanishing gradient problem \citep{glorot_etal_2011}.}

For tile $k$, the output of the network is denoted by
\begin{equation}
\hat{y}_k (n) = y_{1,k}^\text{output}(n),
\end{equation}
where for this study we have a different network for each depth (due to CCA, Sec.~\ref{sec.proposed_method}), and hence only one neuron in the output layer.

In order for the ANN to compute precise solutions, the weights need to be updated iteratively such that 
the following cost function is minimized:
\begin{equation}\label{eq.costFunc}
J(w(n),b(n)) = \frac{1}{2} \sum_{k=1}^{N_\text{title}}  \left( y_k-\hat{y}_k(n) \right) ^2  ,
\end{equation}
 where $y_k$ is the data from the pipeline for tile $k$ (and for a given depth) and $\hat{y}_k(n)$ is the output of the ANN for mini-batch $n$. For the first iteration (mini-batch), the weights are chosen randomly.
The updating of the weights to minimize $J$ is then achieved through back-propagation.

In back-propagation, the layer weight $w_{ji}^{l}(n)$ is adjusted on a layer by layer basis, from the output layer to the first hidden layer. These updates occur for each iteration (mini-batch) $n$, for several passes through all the training data. Each mini-batch consists a number $K(n)$ of tiles. In this work the update of the weights and biases is achieved through stochastic gradient decent:
\begin{equation} 
\begin{aligned}
w_{ji}^{l}(n+1)&=  w_{ji}^{l}(n) -  \eta \frac{\partial J}{\partial w_ {ji}^{l}(n)},\\
b_{j}^l(n+1) &= b_{j}^l(n) - \eta\frac{\partial J}{\partial b_j^l(n)},
\end{aligned}
\end{equation}
where $\eta$ is the the learning rate {which governs how much the weights are changed at each iteration with respect to the cost function}. Here the partial derivatives, or intuitively; the response of the cost function to changes in a specific weight or bias, are computed through 
\begin{equation}
\begin{aligned}
\frac{\partial J}{\partial w_ {ji}^{l}(n)} &= \frac{1}{K(n)}\sum\limits_{k=1}^{K(n)}\delta_{j,k}^l(n)y_{i,k}^{l-1}(n),\\
\frac{\partial J}{\partial b_j^l(n)} & = \frac{1}{K(n)}\sum\limits_{k=1}^{K(n)}\delta^l_{j,k}(n).
\end{aligned}
\end{equation}
where $\delta_{j,k}^l (n)$ is the error of neuron $j$ in layer $l$ for tile $k$ in mini-batch $n$. In order to determine $\delta_{j,k}^l (n)$, one has to know the error of the proceeding neurons. Hence in order to determine all $\delta$, the error of the output neuron is computed first,  
\begin{equation}\label{eq.deltaoutput}
\delta^\text{output}(n) = {y}_{{k}} - \hat{y}_{{k}}(n).
\end{equation}
The errors of the $m_l$ neurons in layer $l$ are then computed from those $m_{l+1}$ neurons in layer $l+1$, through
\begin{equation}\label{eq.deltahidden}
\delta_{j,k}^l(n) =
\mathrm{\varphi}^{\prime}(\upsilon_{j,k}^l(n)) \sum\limits_{q=1}^{m_{l+1}} \delta_{q,k}^{l+1}(n) w_{qj}^{l+1} (n)
\end{equation}
where $\varphi'$ is the derivative of the activation function.

\begin{figure}
\centering

\begin{subfigure}[b]{\linewidth}
\centering
\tikzset{%
  every neuron/.style={
    circle,
    draw,
    minimum size=1cm
  },
  neuron missing/.style={
    draw=none, 
    scale=3,
    text height=0.333cm,
    execute at begin node=\color{black}$\vdots$
  },
}
\begin{tikzpicture}[x=1.3cm, y=1.3cm, >=stealth]

\node  (input) at (0,0.5) {$\delta_{j,k}^l$};

\node [every neuron/.try,label=above:${y}_k-\hat{y}_k(n)$] (hidden) at (3,0.5) {};

\draw [<-] (input) -- (hidden);
\node (text) at (1.5,0.75) {$\varphi'(\upsilon^l_{j,k}(n))$};


\end{tikzpicture}
\caption{}
\end{subfigure}

\begin{subfigure}[b]{\linewidth}
\centering
\tikzset{%
  every neuron/.style={
    circle,
    draw,
    minimum size=1cm
  },
  neuron missing/.style={
    draw=none, 
    scale=3,
    text height=0.2cm,
    execute at begin node=\color{black}$\vdots$
  },
}
\begin{tikzpicture}[x=1.3cm, y=1.3cm, >=stealth]

\foreach \m/\l [count=\y] in {1}
  \node  (input-\m) at (0.1,0.5-\y) {$\delta_{j,k}^l$};

\foreach \m [count=\y] in {1}
  \node [every neuron/.try, neuron \m/.try ] (hidden-\m) at (2,0.5-\y) {$\Sigma$};

\foreach \m [count=\y] in {1,2,missing,3}
  \node [every neuron/.try, neuron \m/.try ] (output-\m) at (5,1.5-\y*1.3) {};



\foreach \l [count=\i] in {1,2,m_{l+1}}
  \node [above] at (output-\i.north) {$\delta_{\l,k}^{l+1}$};

\foreach \i in {1}
  \foreach \j in {1}
    \draw [<-] (input-\i) -- (hidden-\j)
      node [above,midway] {$\varphi'(\upsilon^l_{j,k}(n))$};

\foreach \i in {1}
  \foreach \k [count=\j] in {1,2,m_{l+1}}
    \draw [<-] (hidden-\i) -- (output-\j)
      node [above, midway] {$\quad w_{\k j}^{l+1}$};


\end{tikzpicture}
\caption{}
\end{subfigure}

\caption{Schematics of the Back-Propagation process, where the error of each neuron $\delta$ is computed first from the error in the output layer (a), through all the hidden layers (b). The weights and bias are updated accordingly. Panel (a): Schematic of how the error is computed for the output neuron (Eq.~\ref{eq.deltaoutput}). Note that in this study we use the identity function for the activation function of the output, $\varphi'(\upsilon^l_j(n)) = 1$. Panel (b):
(b) Schematic of how the error of neuron $j$ in hidden layer $l$ is computed from the error of the $m_{l+1}$ neurons in layer $l+1$ (Eq.~\ref{eq.deltahidden}).
}
\label{fig:nn-backprop}
\end{figure}
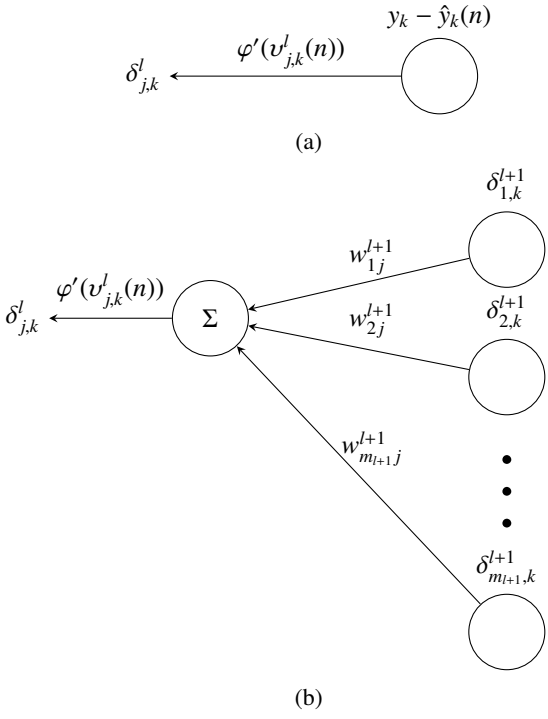

Figure~\ref{fig:nn-backprop} shows a diagram of the Back Propagation process in the both the hidden and output layers.



After updating the layer weights in the backward propagation, the next mini-batch is used in forward and back-propagation to further minimize the cost function. This is repeated (numerous times through the whole dataset) until the maximum allowed number of iterations is reached, or an early stopping criterion is met. The convergence rate of the ANN weights is governed by the learning rate $\eta$, which must be chosen such that the weights converge in a reasonable number of iterations {while still finding the global minimum of the cost function. Typically, the determination of $\eta$ is done experimentally, by slowly increasing $\eta$ until the loss starts to increase. For this study we find ({through} a grid search) $\eta=0.001$ to be a reasonable learning rate.}

In practice a regularization term $\alpha\sum w^2$ is included in Eq.~\ref{eq.costFunc} to penalize complex models that may result in over-fitting. Here we have set $\alpha$ to 0.001.
\section{Performance}\label{sec.performance}
In the previous section, the ANN is predicts flows from the mode shifts given by the ring-diagram pipeline. Here, the performance of the proposed method will be shown and compared to alternative approaches. In this study we use the software packages Python 3.5.4 and {\it scikit-learn} 0.19.1 \citep{scikit-learn}, which are freely available.

\subsection{Experimental Metrics and Setup}
The evaluation metric for this problem will be the root mean square error (RMSE),
\begin{equation}
\textrm{RMSE} = \sqrt{\frac{1}{N_\text{tile}-1} \sum\limits^{N_\text{tile}}_{k=1} (y_k-\hat{y}_k)^2}
\end{equation}
where $y_k$ is the flow for tile $k$ from the ring-diagram pipeline and $\hat{y}_k$ the predicted flow from the machine learning. For this study  the RMSE of each depth will be computed and compared to the mean of the {pipeline inversion} error. A success of an estimator is then measured by how small the prediction's error is compared to the {inverted flow} error. 

In order to verify that the proposed method (preprocessing and ANN) has the ideal performance when compared to other existing methods, an additional evaluation metric is introduced; the computational time (CPU time). As stated previously, the goal of this study is to improve upon the current pipeline, primarily in reducing the computational burden. As such, each step in preprocessing and the chosen ANN architecture will be compared to other methods/architectures in order to demonstrate both the computational burden and accuracy of the proposed method. We note that a large selection of methods exist and a comparison of them all is beyond the scope of this paper. Here we compare to the most common methods.

To avoid overfitting when training supervised methods, the dataset is split randomly into training and testing subsets, in a manner known as {$k$-fold} cross-validation \citep[e.g.][]{mosteller_tukey_1968}. {$k$-fold} cross validation splits the data into {$k$} roughly equal sized sets. The training is then used on $k-1$ subsets and testing (using an evaluation metric) is performed on the remaining subset. 
{This process is applied $k$ times, shifting the testing segment through all of the segments. In doing so the entire data set is used for training and in turn prediction. With each sample in the entire set being used for validation at one time through this process, we can then measure the performance metric of the prediction by averaging the entire set.}
For this study we will use 10-fold cross-validation on the flow data to obtain a complete set of predicted values. These predicted values will be compared with the actual values to obtain the evaluation metric (RMSE) mentioned above.

\subsection{Experimental Analysis}\label{sec.results}


In this section we present the results of each step in the proposed method, using the aforementioned evaluation metrics (computational time and RMSE). For clarity and brevity, we show the comparison results only for a depth of 10.44~Mm. However, the results are consistent with other depths. In order to assess each step in preprocessing, a basic ANN architecture is chosen. The network consists of {1} layer with 10 neurons. In order to assess each step in the preprocessing, the proposed method of Sec.~\ref{sec.proposed_method} is used with only the chosen step varied. The CPU time and root mean square error of the machine learning is determined from the application of $10$-fold cross-validation upon the entire data set (709734 tiles).

\subsubsection{Data Imputation}
Table~\ref{tab:missing} compares different methods for the completion of missing data in the mode fit parameters. In terms of computational gain, it is clear that the mean completion is ideal (2 CPU seconds).


\begin{table}[!t]
\caption{Comparison between methods for the completion of missing data. The bold row is the proposed method.}
\centering
\begin{tabular}{lcccc}
\hline\hline
Method & \multicolumn{2}{c}{CPU Time (s)}  & \multicolumn{2}{c}{Flow RMSE (m/s)} \\ \hline
  & $u_x$&$u_y$ & $u_x$&$u_y$ \\ \hline 
\textbf{Mean}     & \textbf{2}&\textbf{2}&\textbf{8.6}&\textbf{7.3} \\ 
{Median}   &  28&63&8.6&7.3 \\ 
{Most-Frequent} & 31860& 87963 &8.9& 7.4 \\\hline
\end{tabular}
\label{tab:missing}
\end{table}

\subsubsection{Normalization}
Table~\ref{tab:normalization} shows the performance of four different normalization methods, namely, the feature scaling e.g. Minimum-Maximum scaled from 0 to 1 (MM-01) or -1 to 1 (MM-11), Maximum Absolute (MA) and the standardization scaling (SS). The computational burden for each step is rather small, with little difference between them. The same is also true for the RMSE, showing that the choice in normalization is arbitrary for the proposed method.




\begin{table}[!b]
\centering
\caption{Comparison of performance for different methods of normalization. The bold row is the proposed method.}
\begin{tabular}{lcccc}
\hline\hline
Method & \multicolumn{2}{c}{CPU Time (s)}  & \multicolumn{2}{c}{Flow RMSE (m/s)} \\ \hline
  & $u_x$&$u_y$ & $u_x$&$u_y$ \\ \hline 

MM-01 &3&3&8.6&7.3  \\ 
MM-11 &3&4&8.6&7.3 \\ 
MA    &5&3&8.6&7.3 \\  
\textbf{SS} & \textbf{4}&\textbf{4}&\textbf{8.6}&\textbf{7.3} \\ \hline 
\end{tabular}
\label{tab:normalization}
\end{table}

\subsubsection{Feature Selection/Reduction} 
Table~\ref{tab:fs} shows the results of applying different feature reduction methods. The proposed CCA reduction is compared with different feature selection/reduction methods: selecting K best features using f-score \citep{Hand:2001}, applying tree-based methods such as Decision Trees \citep{Hand:2001,Rokach:2014} or Ensemble Trees \citep{Geurts:2006} and Partial Least Squares \citep{Hastie:2001}. The computational times show that our chosen method (CCA) is not the fastest, but when comparing the {RMSE} it out performs the other feature methods. Interestingly, using only a one-component vector achieves good accuracy, and increasing to two component vectors does not improve the result.


\begin{table}[!t]
\caption{Comparison of the performance of different feature selection/reduction methods. The bold row is the proposed method.}
\centering
\begin{tabular}{lcccc}
\hline\hline
Method & \multicolumn{2}{c}{CPU Time (s)}  & \multicolumn{2}{c}{Flow RMSE (m/s)} \\ \hline
  & $u_x$&$u_y$ & $u_x$&$u_y$ \\ \hline 
{Kbest}&6 & 3& 13.41 &11.05 \\ 
{ET} &  44 & 44& 13.85 &8.58\\ 
{DT} & 234 &242 & 13.94 &8.26\\ 
\textbf{CCA-1} & \textbf{14} & \textbf{12} & \textbf{8.61} & \textbf{7.29} \\ 
{CCA-2} & 14 & 14& 8.61 & 7.29\\ 
{PLS} &  9 &11 &13.06 & 10.55\\ \hline
\end{tabular}
\label{tab:fs}
\end{table}

\subsubsection{Neural Networks}
The results thus far have focused on the preprocessing steps of the proposed method. We now focus on the performance of a number of Machine learning techniques and network architectures and compare them to the neural network proposed in this study.

Table \ref{tab:X_ML} compares different supervised \ML{} methods after applying data completion, normalization and feature reduction. The methods examined are Linear Regression, Bayesian Regression \citep{Hastie:2001,Bishop:2006}, Decision Tree \citep{Hand:2001,Rokach:2014}, Ensemble Tree \citep{Geurts:2006}, Random Forest \citep{Breiman:2001}, Gradient Tree Boosting \citep{Friedman:2000}, K-Nearest Neighbor \citep{Hastie:2001} and Support Vector Regression \citep{Bishop:2006,Haykin:2009}. 
The computation time for each method scales with the complexity of the model from the Linear model ($<1$~s) to Support Vector Regression ($\sim50,000$~s). While the proposed ANN takes $200-400$~s for training and predicting, this is still significantly small compared to the current burden of the pipeline.  A comparison of the accuracy shows that the ANN presented in this paper is among the best performing  methods with an RMSE of 8.6~m/s for $u_x$ and 7.3~m/s for $u_y$.

\begin{table}[!b]
\caption{Comparison between supervised ML methods with the same preprocessing. The bold row is the proposed method.}
\centering
\begin{tabular}{lcccc}
\hline\hline
ML & \multicolumn{2}{c}{Training \& Prediction Time} & \multicolumn{2}{c}{RMSE (m/s)}\\ \hline
   & $u_x$&$u_y$  &  $u_x$&$u_y$\\ \hline 
{Lin} &	<1& <1 & 8.8 & 7.3 \\ 
{Bay} & <1& <1 & 8.8 & 7.3 \\ 
{DT}  & 19 & 20 & 8.8 & 7.4\\ 
{ET}  & 2 & 2 & 8.7 & 7.3 \\ 
{RF}  & 39 & 42 &10.5 &9.0 \\ 
{GTB} & 274 & 154 & 8.6 &7.3 \\ 
{KNN} & 50 & 35 & 9.0 &7.7 \\ 
\textbf{NN}& \textbf{170} &\textbf{169} &\textbf{8.6} & \textbf{7.3}\\ 	  
{SVR} & 61441 & 47555 & 8.5 & 7.3 \\ \hline
\end{tabular}
\label{tab:X_ML}
\end{table} 

The previous results have shown that the ANN performs best for the ring-diagram inversions. However, ideal results depend upon the architecture of the ANN, specifically, how many layers and neurons gives the best results for the ANN. 
Table~\ref{tab.nn_layers} shows the performance of the ANN with different numbers of hidden layers. The results show that best performance is obtained with just one hidden layer. The addition of extra layers increases computational burden due to the increase in the complexity of the model. {In terms of how many neurons are required per layer, we have found through experimentation that the RMSE does not improve with more than 10 units.}



\begin{table}[!ht]
\caption{Comparison of ANN performance for different numbers of hidden layers, with 10 neurons in each layer. The bold row is the proposed method.}
\centering
\begin{tabular}{ccccc}
\hline\hline
No. of layers & \multicolumn{2}{c}{CPU Time (s)}  & \multicolumn{2}{c}{Flow RMSE (m/s)}\\\hline
 & $u_x$&$u_y$  &$u_x$&$u_y$\\ \hline 
\textbf{1} & \textbf{170}  & \textbf{169}& \textbf{8.59} & \textbf{7.29} \\
{2} & {360}  & {194}& {8.59} & {7.29} \\  
3 & 342  & 330 & 8.59 & 7.30 \\  
4 &  427 & 439 & 8.61 & 7.30 \\  
5 &  3099 & 1695 & 8.62 & 7.30 \\  
\hline
\end{tabular}
\label{tab.nn_layers}
\end{table}


\subsubsection{RMSE of Model with Depth}
The results thus far have been shown for one depth (10.44~Mm), Fig.~\ref{fig:depth_RMSE} shows the differences between mean {inversion} error and the RMSE of the ANN for all depths below the photosphere. We have ignored the results at $z_t=0$~Mm (photosphere) due to inconsistency in the inverted flows in the HMI pipeline. The results of Fig.~\ref{fig:depth_RMSE} show that proposed ANN of this study achieves a RMSE that is generally below the {inversion} errors reported in the pipeline. While the errors are not directly comparable, the results provide confidence in the results of the \ML{}. 
Additionally, while the errors reported in the pipeline are similar for $u_x$ and $u_y$, there is a difference in the \ML{} errors. This is due to errors in the \ML{} resulting from a different variances in $u_x$ and $u_y$ with the former having larger variance than the latter. This variance leads to a more difficult fit for the model, thus higher error. Additionally, the machine learning may have difficulties fitting $u_x$ due to systematics in the patch tracking in the $x$ direction. 
\subsubsection{{Effect} of realization noise}
\begin{figure}
    \centering
    \includegraphics[width=\linewidth]{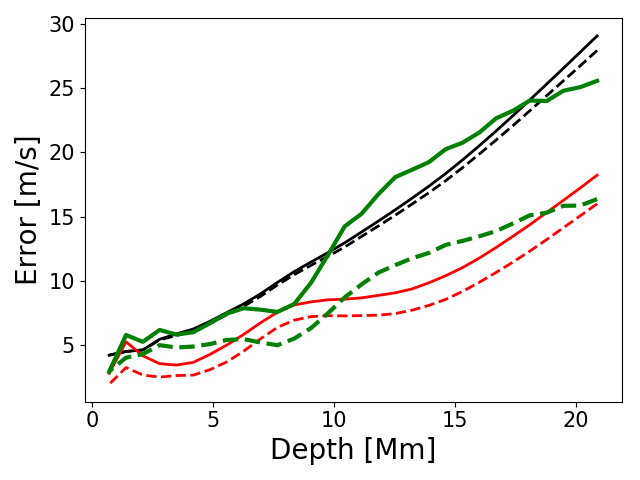}
\caption{{Comparison of the pipeline inversion error (black lines) and \ML{} error (RMSE, red lines), for $u_x$ (solid) and $u_y$ (dashed) for all depths below the surface. The green lines are the standard deviation of the prediction values after noise realizations are added to the mode fits.} }
\label{fig:depth_RMSE}
\end{figure}

{
So far, we have used just the mode fits in predicting the flow values. However, the determination of these mode fits is not exact and hence the {effect} of mode fit error on the \ML{} model must be considered.
In order to examine the propagation of errors we take CR~2107-2201 and build the ANN described previously. {The mode fits of the remaining 10 rotations} are then perturbed by a realization of the errors. Predictions are then made with the new noisy data. This process is repeated 1000 times and the deviation computed. Figure~\ref{fig:depth_RMSE} shows the averaged deviation as a function of depth given the noisy data. For $u_x$ the errors are of the order of the inversion pipeline, while the $u_y$ are less. This result is not unexpected. The ANN was trained to find a particular relationship between the mode fits and the flows across the disk. By adding errors to the data, these errors propagate through the model, producing a deviation in the predictions higher then the RMSE of the model. The fact {that} the deviations are not significantly greater {than} those reported in the pipeline is a good indicator of the robustness of the model to errors.  
}

\subsubsection{RMSE vs. Number of Samples}

\begin{figure}
\includegraphics[width=\linewidth]{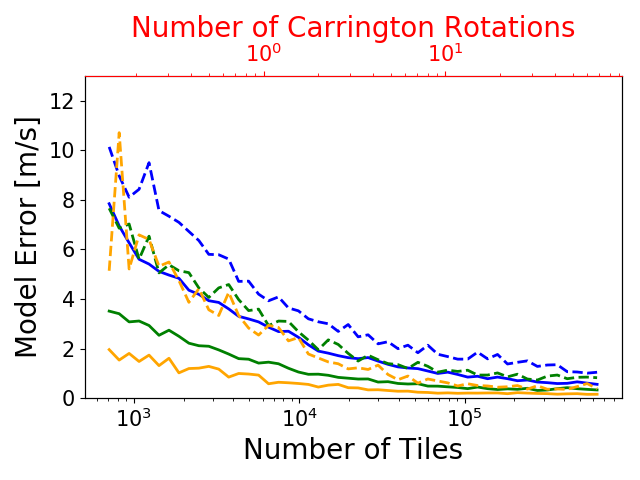}
\caption{{The model error (RMSE) between an ANN trained with the complete 105 CRs and one trained with a subset (using k-fold cross-validation). The errors converge for both $u_x$ (solid) and $u_y$ (dashed) as the subsample size approaches the full sample size.} Here we only show the results for depths of 20.88~Mm (blue), 10.44~Mm (green) and 0.696~Mm (orange).}
\label{fig.training_number}
\end{figure}

We conclude this section by addressing the question of how many samples are needed for an accurate ANN. In the field of machine learning, the answer to the common question of how to get a better model is; more data. Thus, we compute the model error (RMSE) between a model that is trained with the complete 105 Carrington rotations, and those trained with only a subset. Again we use 10-fold cross-validation in prediction. Figure~\ref{fig.training_number} shows the convergence of models trained with a increasing subset size to that trained with the full data set. The results show that with just a small subset of around 1-10CRs ($\sim6000$-60,000 Tiles) the model error between the predictions of an ANN trained with the 105~CRs and an ANN trained with a subset is below {4}~m/s for the three depths examined. The results also show that $u_y$ converge slower than $u_x$. Larger sample sizes slowly improve the model {to that of the full set.}

\section{Case Study: Equatorial Rossby Waves}\label{sec.rossby}
\begin{figure}
\includegraphics[width=\linewidth]{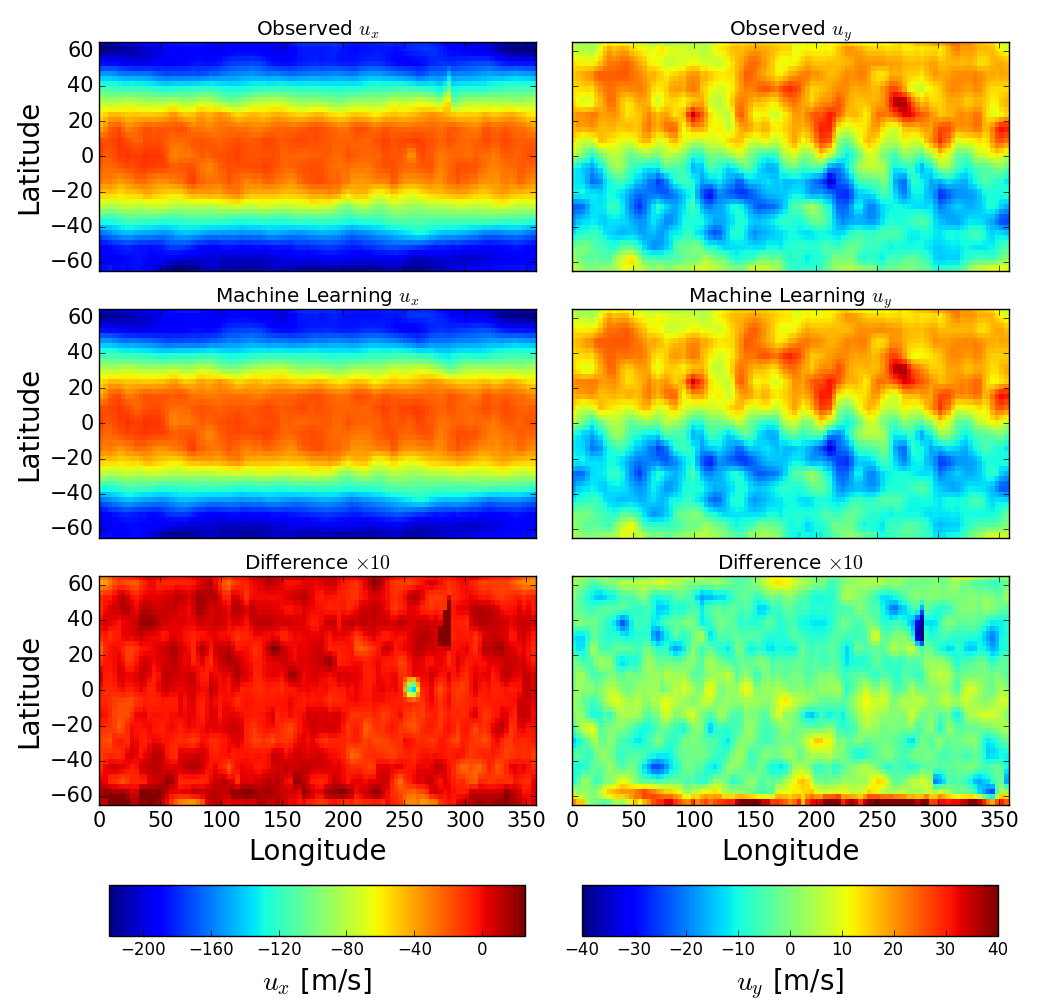}
\caption{Comparison of the flows maps for $u_x$ (left) and $u_y$ (right) between the {pipeline} (top), machine learning (middle) and the difference (bottom, scaled by factor 10) {at a target depth of 0.696~Mm}. The maps are generated from a time average over the Carrington rotation 2100. {The predicted values were obtained from an ANN trained using CR~2107-2201.}}
\label{fig.flow_comp}
\end{figure}

In fitting any model to a vast amount of data, there is a possibility that the subtle helioseismic features in each tile are removed or altered. Figure~\ref{fig.flow_comp} shows the {pipeline} and machine learning flows for both components of the flow, and their difference. Close examination shows that while the general structure is nearly identical, some small features present in the {pipeline flows} are not present in the machine learning (e.g. at longitude and latitude $280^\circ$ and $40^\circ$, respectively). While these features appear as artifacts to the keen observer, it raises the question of whether the \ML{} model (in an effort to fit a model) overlooks helioseismic signatures seen after averaging over long time scales. To explore this possibility, we examine equatorial Rossby waves in the same manner as \citet{loeptien_etal_2018}, but with the predicted flows.

\citet{loeptien_etal_2018} recently reported the unambiguous detection of retrograde-propagating vorticity waves in the near-surface layers of the Sun. These waves exhibit the dispersion relation of sectoral Rossby waves. Solar Rossby waves have long periods 
of several months with amplitudes of a few meters per second, making them difficult to detect without averaging large amounts of data. To detect these Rossby waves in both the raw data and the ML data, we follow the technique of \citet{loeptien_etal_2018}, specifically:
\begin{enumerate}
\item Flow tiles ($u_x,u_y$) are sorted into cubes of Latitude, Stoney-Hurst Longitude and time
\item The one year-frequency signal (B-angle) is removed
\item Missing data on the disk are interpolated in both time and latitude
\item Data exceeding a distance of $65^\circ$ from disk center are neglected
\item The data is remapped into a reference frame that rotates at the equatorial rotation rate (453.1~nHz). Then transformed back to a Carrington longitude grid
\item The longitude mean is subtracted
\item The vorticity is computed
\item Spherical harmonic transforms (with $m=\ell$) and temporal Fourier transforms are applied to produce a power spectrum
\end{enumerate}
We apply this procedure to the flow maps at a depth of 0.696~Mm and 20.88~Mm.

\begin{figure}
\includegraphics[width=\linewidth]{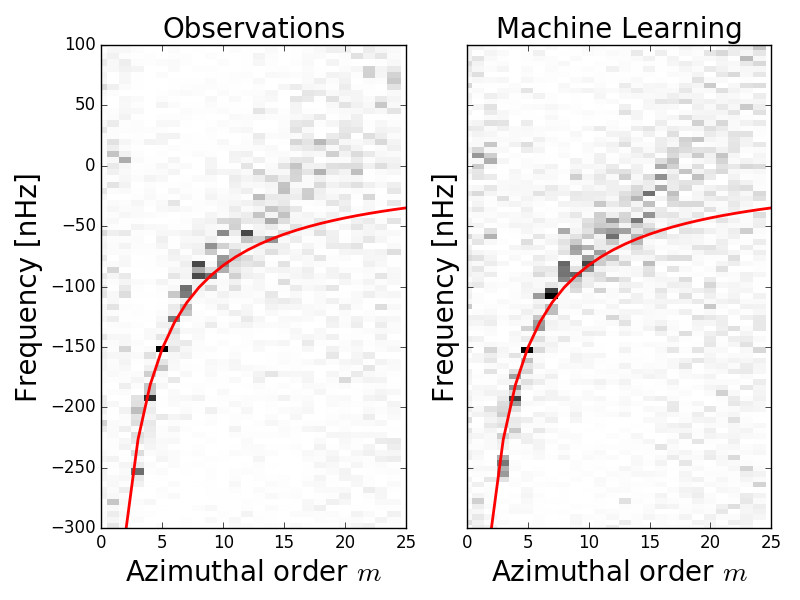}
\caption{Comparison of the Rossby wave power spectrum computed from $\sim6$~years ({CR~2089-2201}) of {pipeline} data (left) and from the machine learning (right) for a depth of 0.696~Mm. The ML method was trained using 10 Carrington rotations {(CR~2079-2088)}. The red line is the dispersion relation $\omega=-2\Omega_\textrm{ref}/(m+1)$ of sectoral Rossby waves, measured in a frame rotating at the equatorial rotation rate $\Omega_\textrm{ref} = 453.1$~nHz.}
\label{fig.rossbyWaveComp}
\end{figure}

\begin{figure}
\includegraphics[width=\linewidth]{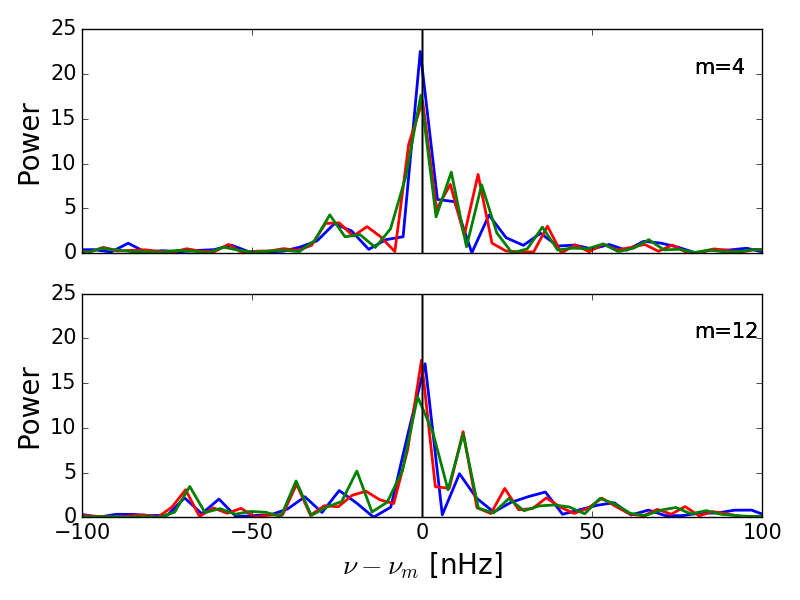}
\caption{Comparison of sectoral power spectra at a depth of $20.88$~Mm for the Rossby waves with azimuthal orders $m=4$ and $m=12$. The results include the HMI observations (blue) and machine learning trained using 1 {(CR~2079)} and 10 {(CR~2079-2088)} Carrington rotations (red and green, respectively). For each $m$, the power spectrum frequency ($\nu$)  is centered on the Rossby wave frequencies ($\nu_m$) reported by \cite{loeptien_etal_2018}.}
\label{fig.rossbyWaveComp_slices}
\end{figure}

In order to examine if Rossby waves are still present in the \ML{} we take the results from the ANN model outlined in the previous sections for CR~2097-2181. The Rossby wave procedure outlined above was then followed using these new maps. Figure~\ref{fig.rossbyWaveComp} compares the Rossby wave power spectrum from the {pipeline flows} and the ML, computed at a depth of 0.696~Mm. By visual inspection they are very similar, validating the ML method's ability to recover the presence of Rossby waves. Additionally, Fig~\ref{fig.rossbyWaveComp_slices} shows slices of the power spectrum for different azimuthal order $m$, for the proposed method trained with 1, 10 and 20 Carrington rotations (from the unused CR~2182-2201). The results show that just using as small a sample as 1 Carrington rotation ($\sim 6800$ tiles) for training the ML model, can produce a model that captures the Rossby wave power spectrum reasonably well.

\section{Conclusion}\label{sec.conclusions}
Local helioseismology has a plethora of raw observed data of the Sun. Despite 50 years of observations and analysis, we are still have no consistent and complete answer to the Sun's internal structure. The computational field of machine learning and artificial intelligence has grown in both usage and capabilities in the last few decades, and has shown promise in other fields in ways that could be extended to local heliseismology.

In this study we have shown that \ML{} provides an alternative to computationally expensive methodologies. We have shown that in utilizing the data of 8 years of HMI observations, we can use an ANN model to predict future flow data with an {RMSE} that is well below that of the observations, while maintaining the flow components of interest to local helioseismology. {Additionally, we find that the propagation of noise realization results in a deviation of the flow values that is of the order of the pipeline errors.} The computational burden was previously 31 CPU hours for 1 Carrington rotation worth of data. With a trained ANN the computational costs is now $10^{-3}$ CPU hours. While we have focused our efforts in obtaining an accurate ANN model, the results of Sec.~\ref{sec.performance} show that any number of common architectures or preprocessing can obtain a reasonable model for future predictions. {Yet,} non-linear models (such as the proposed ANN here) can capture some of the non-linearity (e.g. noise or missing modes) that occurs between all tiles across the disk. 

Here we have only shown the computational efficiency gain for through the application of \ML{}, but future improvements can be made. The method presented here has been purely data driven, without introducing constraints a-priori. Recent studies \citep[e.g.][]{raissi_etal_2017a,raissi_etal_2017b} have shown that physics informed neural networks can be built, capable of enforcing physical laws (e.g. mass conservation) when determining the \ML{} model. While the constraint of physical laws is beyond the scope of the work here, these studies demonstrate the capabilities of applying \ML{} in determining subsurface solar structures, to which we have prior knowledge of constraints. Additionally, the use of synthetic ring diagrams computed using computational methods with \ML{} could improve current capabilities of the pipeline in probing solar subsurface flows. Thus the application of machine learning and deep learning techniques present a step forward for local helioseismic studies.
\begin{acknowledgements}\\
This work was supported by NYUAD Institute Grant G1502 ``NYUAD Center for Space Science''. RA acknowledges support from the NYUAD Kawadar Research Program.  The HMI data is courtesy of NASA/SDO and the HMI Science Team. Computational resources were provided by the German Data Center for SDO funded by the German Aerospace Center. SMH and LG thank the Max Planck partner group program for supporting this work.\\
{Author Contributions:} CSH and LG designed the research goals. RA proposed the machine learning methodology for predicting flows and performed experimental and computational analysis. CSH prepared data and performed helioseismic analysis. All authors contributed to the final manuscript. We thank Aaron Birch and Bastian Proxauf for helpful insight on the ring-diagram pipeline.
\end{acknowledgements}

\bibliographystyle{aa}
\bibliography{biblio}

\end{document}